# A rigorous link performance and measurement uncertainty assessment for MIMO OTA characterisation

*Min Wang\*†, Tian-Hong Loh†, David Cheadle†, Yongjiu Zhao\*, Yonggang Zhou\**

\*Nanjing University of Aeronautics and Astronautics, Nanjing, Jiangsu,  211106,  China, wangmin_win@126.com
†National Physical Laboratory, Hampton Road, Teddington, Middlesex TW11 0LW, UK, tian.loh@npl.co.uk

## Abstract

In this paper, a rigorous link performance test for MIMO OTA (multiple-input-multiple-output over-the-air) characterisation with traceable signal-to-interference-plus-noise ratio (SINR) is presented. Measurements were made with three different testbeds which represent the 4G (fourth-generation) LTE (long term evolution) SISO (single-input-single-output), 4G LTE MIMO and 5G (fifth-generation) millimetre-wave (mm-wave) MIMO communication systems, respectively, in the small antenna radiated test (SMART) screened fully anechoic chamber, screened control room and reverberation chamber at the UK National Physical Laboratory (NPL). The measurement campaign comprised of automated data acquisition of channel power, downlink (DL) & uplink (UL) error vector magnitude (EVM) and throughput. The measurement repeatability has been assessed with standard deviation plotted as error bars and the uncertainty sources in the OTA test are analysed.

## 1  Introduction

Multiple-input-multiple-output (MIMO) antenna systems (see Figure 1) play a significant role in 4G (fourth-generation) and 5G (firth-generation) communications [1, 2]. Transmission diversity, such as spatial and polarisation diversity, is key features of MIMO communication systems that used to maximise the available throughput to a single or multiple users. The use of broadened spectrum in new 5G wireless technologies alongside legacy 4G systems may incur interferences from adjacent band. Unpredicted interferences may also be generated from non-desired paths within the system or from other systems. In order to understand the end-to-end reception performance of a wireless device, over-the-air (OTA) testing is needed. Rigorous MIMO OTA characterisation is challenging because of its complexity [3]. In particular, uncertainty contribution arise due to multipath. Wireless industry groups such as 3GPP (Third Generation Partnership Project) [4] has spent fruitful efforts on the development of MIMO OTA standardisation.

Signal-to-interference-plus-noise ratio (SINR) is a quantity which is widely used in theoretical studies of channel capacity in wireless communications and sets an upper bound on the information carrying capacity of a communications system [5]. EVM is a quality parameter of digital modulation and demodulation. The Linear relationship between SINR and root-mean-square (RMS) error vector magnitude (EVM), in a multi-user scenario has been investigated in [6].

In this paper, a series of link performance OTA tests with traceable SINR are presented to further assess the suitability of linear EVM-SINR relationship for SISO and MIMO communications in different electromagnetic environments. The fundamental theory and uncertainty analysis of SINR and EVM are introduced in Section 2. The measurement campaign was carried out by considering three different scenarios, namely SISO communication in the anechoic chamber (AC), 4G LTE 2 by 2 MIMO communication system with the 4th transmitting mode (TM4, close-loop spatial diversity) in a reverberation chamber (RC) and the NPL 5G mm-wave MIMO testbed with transmitter diversity technology in the screened control room. All the measurements are automated for data acquisition of channel power, error vector magnitude (EVM), and downlink (DL) & uplink (UL) throughput. The relationship between SINR and EVM is investigated with respect to different transmitting configurations. The measurement repeatability was assessed with uncertainty scales plotted as error bars. All the channel power measurements are referenced to a traceable spectrum analyser and the results have been calibrated and corrected where the associated measurement uncertainties have been taken into account for obtaining traceable SINR.

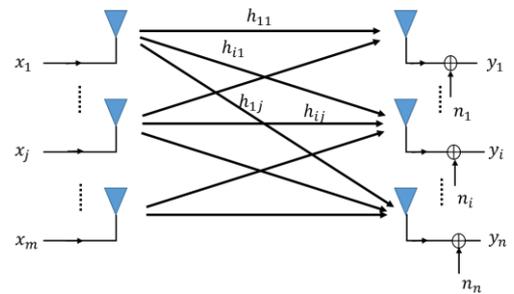

Figure 1: Diagram of MIMO system

## 2  Theory

### 2.1 Definitions of SINR and RMS EVM

The standard definition of SINR for in-band signal interferences is shown in Equation (1) [7]:

$$\text{SINR} = \frac{P_{tS}|H_w(t)|^2}{\sum_{i=1}^{N_I} P_{tI}^i |H_i(t)|^2 + \sigma_n^2} \qquad (1)$$
$$= \frac{P_{rS}}{\sum_{i=1}^{N_I} P_{rI}^i + \sigma_n^2}$$

where $P_{tS}$ and $P_{rS}$ are the signal source power and receiving power respectively, $H_w(t)$ represents the propagation channel of the desired signal, $P_{tI}^i$ and $P_{rI}^i$ are the $i^{th}$ interference source power and receiving power respectively, $H_i(t)$ is the propagation channel of each interference signal, $N_I$ is the total number of interferences and $\sigma_n^2$ is the mean power of the noise floor.

In this work, traceable SINR was measured at the receiving antenna port based on the sampled digital signals captured by vector signal analyser, and can be calculated as:

$$\text{SINR} = \frac{\frac{\sum_i^{N_{symbol}} |S_i|^2}{N_{symbol}}}{\sum_{j=1}^{N_{intf}} \frac{\sum_i^{N_{symbol}} |I_{ji}|^2}{N_{symbol}} + \sigma_n^2} \qquad (2)$$

where $S_i$ is the $i^{th}$ desired signal symbol, $I_{ji}$ is the $i^{th}$ interference signal symbol of the $j^{th}$ interference source.

The RMS EVM is most commonly used for evaluating EVM, and their definition are shown in Equation (3) to (7) [8]:

$$\text{EVM} = \frac{|s_{act} - s_{ref}|}{|R_i|} \times 100\% \qquad (3)$$

$$\text{MagErr}_{rms} = \sqrt{\frac{\sum_{i=1}^{N_{symbol}} (|s_{act}| - |s_{ref}|)^2}{\sum_{i=1}^{N_{symbol}} |s_{ref}|^2}} \times 100\% \qquad (4)$$

$$\text{PhaseErr}_{rms} = \sqrt{\frac{\sum_{i=0}^{N_{symbol}} (args_{act} - args_{ref})}{N_{symbol}}} \qquad (5)$$

$$\text{EVM}_{rms} = \sqrt{\frac{1}{N_{symbol}} \cdot \frac{\sum_{i=1}^{N_{symbol}} (|s_{act} - s_{ref}|)^2}{\sum_{i=1}^{N_{symbol}} |s_{ref}|^2}} \times 100\% \qquad (6)$$

$$\text{Normalised EVM}_{rms} = \sqrt{\frac{\sum_{i=1}^{N_{symbol}} (|s_{act} - s_{ref}|)^2}{\sum_{i=1}^{N_{symbol}} |s_{ref}|^2}} \times 100\% \qquad (7)$$

where $s_{act}$ is the received modulated symbol vector, $s_{ref}$ is the reference modulated symbol vector. The assumption in this work is that the interference signal has Gaussian like distribution, which was implemented as Gaussian White Noise (GWN) in software and filtered by a digital bandpass filter. This allows a direct link between the error vectors of the received waveform signals and the EVM of demodulated symbols.

### 2.2 Interferences from the same system

One of the key technologies for 5G communications is massive MIMO multi-beamforming. Three typical algorithms for beamforming, namely max ratio (MR), zero forcing (ZF) and minimum mean square error (MMSE). These algorithms apply precoding based on channel-state information (CSI) to control the relative phase and amplitude of the signal at each transmitter, in order to create a pattern of constructive and destructive interference. With the uplink communication, pilot signals are sent to the base station from all mobiles to gain channel state estimation such that they could apply an optimal pre-coder to reach multiple users. For any single receiving antenna,

$$y_i = \sum_{j=1}^{N_S} h_{ij} x_j w_j + \sum_{k \neq j}^{N_{intf}} h_{ik} x_k w_k + n_i, i = 1, \cdots, N_R \qquad (8)$$

where $N_S$ is the number of signal sources, $w_j$ is the linear pre-coder. It is worth noting that pre-coders from neighbouring cells or a high-speed mobile scenario will cause inter-cell interference. Since the signal and interference are generated by the same base-station system, both the signal and interference channel coefficients can be entirely estimated. Due to the linearity of discrete Fourier transform (DFT) process, for a length $N$ input waveform vector $Y$, take the OFDM waveform as an example, the demodulated signal is a length $N$ vector y, with elements

$$y(k) = \sum_{n=1}^{N} Y(n) * exp\left(\frac{-j*2\pi*(k-1)*(n-1)}{N}\right), 1 \leq k \leq N \qquad (9)$$

Introducing the interference signals, the demodulated symbol obtained becomes:

$$\begin{aligned} y'(k) &= y(k) + \sum_{h=1}^{N_{intf}} \sum_{n=1}^{N} \Delta Y_h(n) \\ &\quad * exp\left(\frac{-j*2\pi*(k-1)*(n-1)}{N}\right) \\ &= y_i(k) + \sum_{h=1}^{N_{intf}} i_h(k) \end{aligned} \qquad (10)$$

Then the SINR for each demodulated symbol can be derived:

$$\text{SINR} = \frac{|y_i(k)|^2}{\sum_{h=1}^{N_{intf}} |i_h(k)|^2 + \sigma_n^2} \qquad (11)$$

Thus a relationship has been found between the linear SINR and the EVM as follows [7]:

$$\text{EVM}_{rms}(\%) = \frac{A}{\sqrt{SINR}} \qquad (12)$$

### 2.3 Interferences from the different systems

When interferences come from different system the relevant analysis will be much more complicated since the mobile terminal cannot identify the number and channel coefficients of the interference signals, which results in incorrect channel estimation for the considered system and additional uncertainty in the demodulated symbols, this idea will present an issue for OTA testing. For a transmitting diversity scheme with no interference, one assumes the adjacent two symbols sent by the two transmitting antennas are:

$$x^1 = [x_1, -x^*_2] \quad (13)$$
$$x^2 = [x_2, x^*_1]$$

Through the propagation channel, the received symbols over two symbol intervals can be written as:

$$\begin{bmatrix} y(f_1) \\ y^*(f_2) \end{bmatrix} = \begin{bmatrix} h_{11} & h_{12} \\ h^*_{12} & -h^*_{11} \end{bmatrix} \begin{bmatrix} x_1 \\ x_2 \end{bmatrix} + \begin{bmatrix} n_1 \\ n_2 \end{bmatrix} = H \begin{bmatrix} x_1 \\ x_2 \end{bmatrix} + n \quad (14)$$

Then the symbols are extracted based on the propagation channel inform of H-matrix as follow, $*^H$ refers to the conjugate operation:

$$r = \begin{bmatrix} r_1 \\ r_2 \end{bmatrix} = H^H \cdot \left( H \begin{bmatrix} x_1 \\ x_2 \end{bmatrix} + n \right)$$
$$= \begin{bmatrix} (|h_{11}|^2 + |h_{12}|^2)x_1 + h^*_{11}n_1 + h_{12}n_1^* \\ (|h_{11}|^2 + |h_{12}|^2)x_2 - h^*_{11}n_2^* + h^*_{12}n_0 \end{bmatrix} \quad (15)$$

Following introduction of interference signal, Eqs (14) and (15) become

$$\begin{bmatrix} y'(f_1) \\ y'^*(f_2) \end{bmatrix} = \begin{bmatrix} h_{11} & h_{12} \\ h^*_{12} & -h^*_{11} \end{bmatrix} \begin{bmatrix} x_1 \\ x_2 \end{bmatrix} + \begin{bmatrix} i_1 \\ i_2 \end{bmatrix} + \begin{bmatrix} n_1 \\ n_2 \end{bmatrix}$$
$$= \begin{bmatrix} h'_{11} & h'_{12} \\ h'^*_{12} & -h'^*_{11} \end{bmatrix} \begin{bmatrix} x_1 + i_1 \\ x_2 + i_2 \end{bmatrix} + \begin{bmatrix} n_1 \\ n_2 \end{bmatrix} = H'(x + I) + n \quad (16)$$

$$r' = \begin{bmatrix} r'_1 \\ r'_2 \end{bmatrix} = H'^H \cdot \left( H \begin{bmatrix} x_1 \\ x_2 \end{bmatrix} + \begin{bmatrix} i_1 \\ i_2 \end{bmatrix} + \begin{bmatrix} n_1 \\ n_2 \end{bmatrix} \right) \quad (17)$$

When the pilot symbols are contaminated by the interference symbols, the deviated H-matrix will bring errors and more uncertainties to the MIMO decoding.

## 3 Experiments

### 3.1 Sub-6G band BSE-UE test for uplink performance

The 4G LTE SISO and MIMO measurements were made, respectively, in two different environment, namely, the NPL SMART chamber and its screened control room. The chamber is a fully anechoic chamber with dimensions of 7.15 m × 6.25 m × 6.25 m. The chamber operates at frequencies above 400 MHz and is temperature controlled at 23°C ± 2°C. The control room is a screened room with metallic walls and equipment. In this work, it is been considered as a 'real world' multipath environment as compared with a fully anechoic chamber (i.e. RF reflectionless environment). The base station emulator (BSE) employed is Keysight E6621A PXT wireless communications test set and the user equipment (UE) employed is NETGEAR MR1100 Mobile Router. The BSE enables testing and analysing of the link performance and signalling power level of the UE based on the 3GPP standard. A 45 degree slant MIMO antenna is employed at the UE and an ETS-Lindgren 3117 double-ridge horn antenna is employed at the BSE.

During the measurement in SMART chamber, the UE was positioned on the Kevlar tower. During the measurement in the control room, the UE was positioned on the test bench. The measurement setup in the SMART chamber is shown in Figure 2. The signal power level (RF1) is swept from 0 dBm to -60 dBm in 1 dBm steps, and at each power level, the measurement repeats 100 times. Note that the link becomes unstable when RF1 <= -50 dBm. The measurement repeatability has been assessed with standard deviation plotted as error bar. As depicted in Figure 2(b), the standard deviation of uplink EVM increases with the increase of DL RF power together with the non-linearity issue between −10 dBm and 0 dBm.

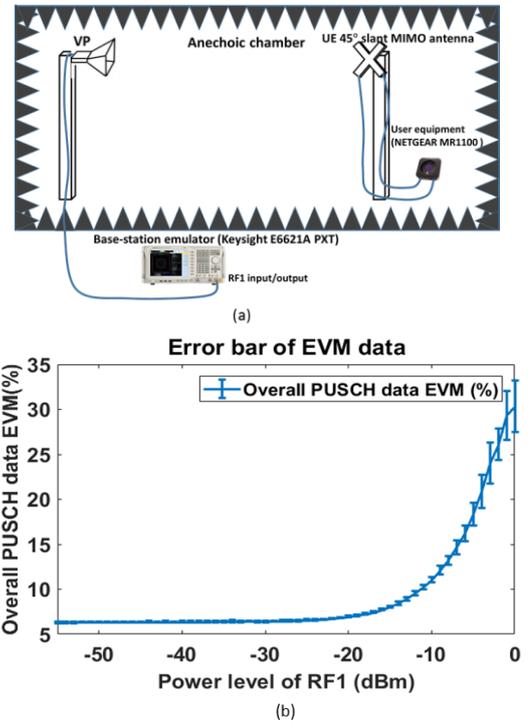

Figure 2: (a) Diagram of LTE SISO testbed setup in anechoic chamber; (b) Overall uplink data EVM with swept downlink signal power level.

Further measurements were carried out with a 2 by 2 MIMO setup with a single interference source in the reverberation chamber. The setup is shown in Figure 3. The BSE chose the closed-loop spatial multiplexing transmitting scheme and the UE configuration was set as 'all up' to maintain a stable output power. An individual interference signal was generated by a vector signal generator. For channel power measurement, all the channel power measurements are calibrated to the antenna port plane referenced to a traceable spectrum analyser (Agilent E4440A PSA) where the losses and mismatch introduced by the coupler & its side-arm, cable, PSA have been taken into account for obtaining traceable SINR. In addition to the repeatability standard deviation, its measurement uncertainty for $k = 2$ (i.e. 95% confidence interval) has also been assessed and plotted as error bar. The following presents the formula for calculating the channel power measurement relative uncertainty, $U_{ChPow}$:

$$U_{ChPow} = 10\log_{10}\left(1 + \frac{2\sigma}{\mu}\right) + U_{FreResp} + U_{InputAtt} + U_{Abs} \quad (18)$$
$$+ U_{RBW} + U_{InputMixer}$$

where

$\sigma$ is the repeatability standard deviation
$\mu$ is the average
$U_{FreResp} = 0.38$ dB, is the frequency response uncertainty
$U_{InputAtt} = 0.2$ dB, is the input attenuation switching uncertainty
$U_{Abs} = 0.24$ dB, is the absolute amplitude accuracy
$U_{RBW} = 0.03$ dB, is the resolution bandwidth switching uncertainty
$U_{InputMixer} = 0.07$ dB, is the input mixer level linearity

As shown in the results in Figure 4 to Figure 6, the uplink EVM increases with the increase in uplink SINR. This measurement results demonstrates that the downlink signals, due to possible non-linearity and non-ideal duplexer isolation of the receiver, act as an interference source to the uplink signal. From the measured downlink throughput result, one observes that the downlink MIMO communication in a multipath environment stop sending data at a cut-off of DL SINR <= 0 (i.e. at about the interference signal level).

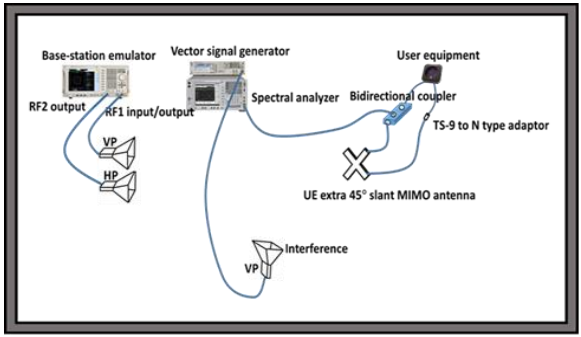
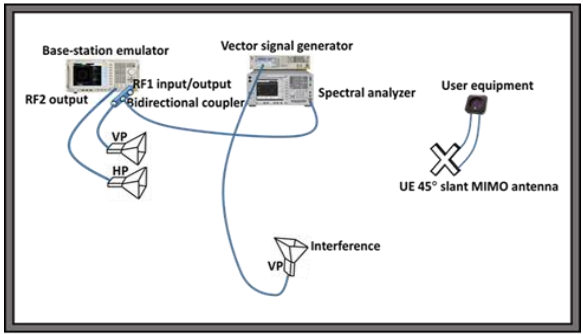

Figure 3: Diagram of LTE MIMO uplink performance test: (a) swept uplink SINR; (b) swept downlink SINR.

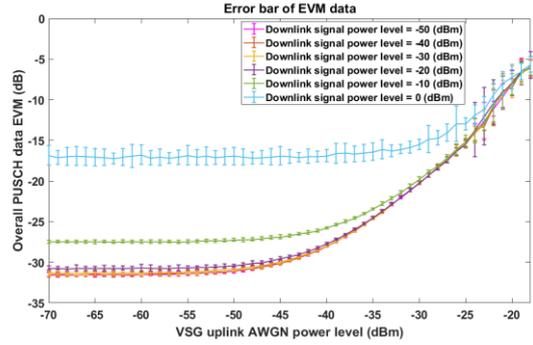

Figure 4: Uplink SINR-EVM interrelation with different downlink signal levels.

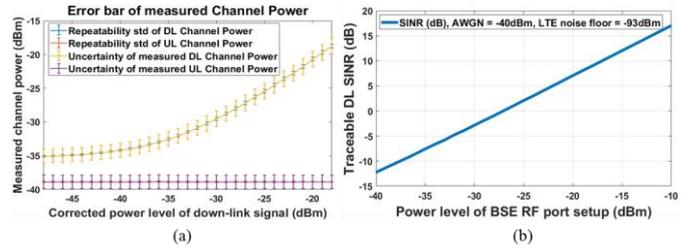

Figure 5: Channel power measurement results with 200 repeats and downlink AWGN interference signal at -40 dBm on both RF ports: (a) downlink and uplink channel power; (b) Traceable downlink SINR.

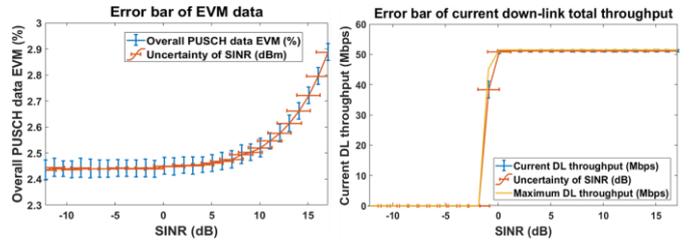

Figure 6: (a) Uplink EVM measurement results; (b) Downlink throughput measurement results

### 3.2 5G testbed measurements

Similar work has been carried out with the NPL 5G millimetre-wave software defined radio testbed, which is a configurable mm-wave MIMO testbed and is capable of performing spatial diversity transmission. The whole system includes four vector signal transceiver (NI PXIe 5644R VST modules) system modules with a real-time signal processing software defined radio (SDR) capability, two pairs of standard gain horns at the transmit and receive ends and the frequency up and down conversion hardware (see Figure 7). This was setup to operate at a centre frequency of 26 GHz with 20MHz bandwidth. Cyclic prefix orthogonal frequency division multiplexing (CP-OFDM) signal generation and measurement is performed using a pair of sub-6 GHz VSTs. The software to control the NPL 5G testbed is written in LabVIEW, and Matlab is used for signal generation and processing. SINR and RMS EVM are calculated for each sub-frame.

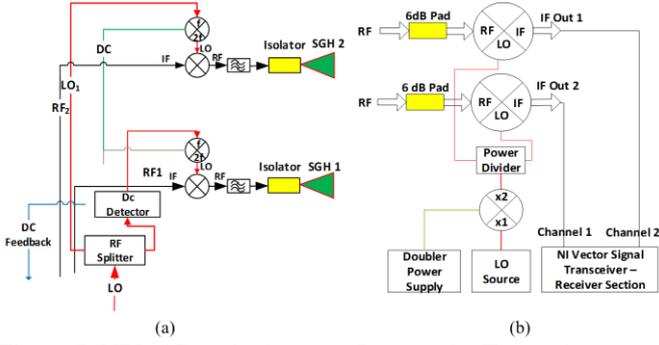

Figure 7: NPL 5G testbed system Layout: (a) Transmit system; (b) Receive system.

At first, the channel power is calculated for each sub-frame from the measured IQ data captured by the vector signal analyser (VSA) at the receiver. The channel measurement bandwidth is the same as the expected occupied bandwidth of the signal. The channel power is used to determine signal-to-interference-plus-noise ratio (SINR). The SINR has been calibrated referenced to measurements made using a traceable spectrum analyser.

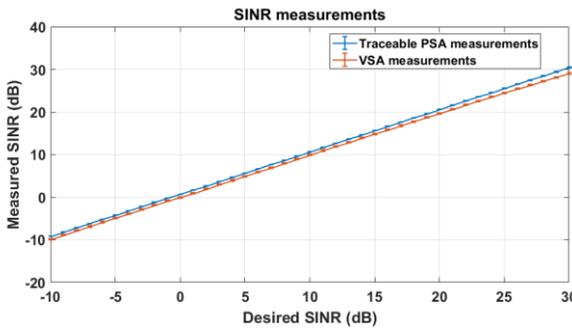

Figure 8: Comparison of SINR measured by PSA and VSA.

Initially, the system setup was a 2 by 1 MISO using transmission diversity mode, the interference signals were in-band GWN generated by other VST modules. The MISO decoder used the measured IQ data at the receivers and previously obtained channel H-matrix to recover two simultaneously transmitted sub-frames, received interference power level was adjusted at the receiving end to obtain a controllable SINR. Equations (6) and (7) were used to calculate un-normalised and normalised RMS EVM respectively, as shown in Figure 9 and Figure 10. One observes that there's a perfect linear relationship between SINR and RMS EVM, in addition, the value of the gradient, determined by A is highly dependent on the quadrature amplitude modulation (QAM) order when the RMS EVM is presented with a non-normalised value, on the contrary, using normalised EMS EVM with respect to the number of symbols, a constant gradient $A$ is obtained for different modulation schemes. As the interferences here are generated by filtered GWN, so no matter the number of interferers, the total interference would appear to just cause an EVM as though it was an increase in WGN component.

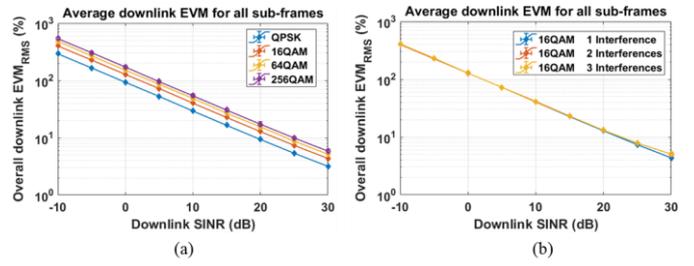

Figure 9: Un-normalised RMS EVM of the receiving signal interfered by in-band GWN demodulated using previously obtained channel H-matrix and controllable SINR at the receiving end: (a) comparisons of different symbol modulation schemes, A = [93.3 128.6 153 172.5]; (b) comparisons of different numbers of interferences.

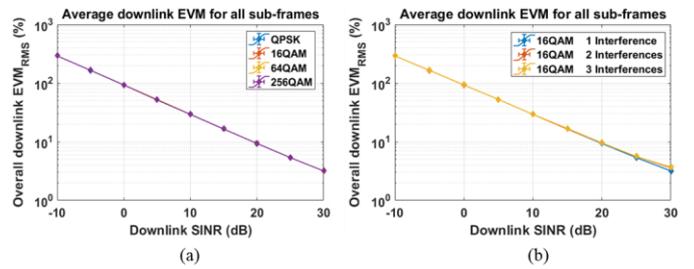

Figure 10: Normalised RMS EVM of the receiving signal interfered by in-band GWN demodulated using previously obtained channel H-matrix and controllable SINR at the receiving end: (a) comparison of different symbol modulation schemes, A = 93.3; (b) comparison of different numbers of interferences.

Further measurements were taken with in-band normal distributed LTE CP-OFDM signals as interferences, random frame offset was applied to the interference waveform. From the results shown in Figure 11 and Figure 12, more uncertainty are observed due to inhomogeneous in-band power spectral density (PDF) of the interference signal.

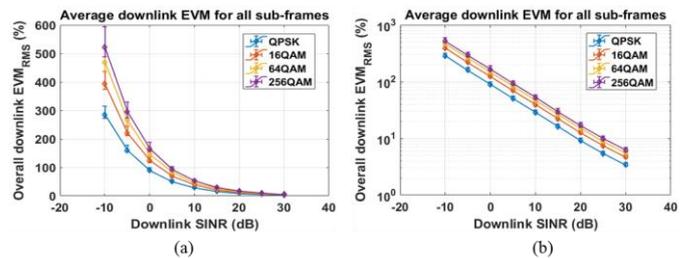

Figure 11: Un-normalised RMS EVM of the receiving signal interfered by in-band LTE CP-OFDM signals demodulated using previously obtained channel H-matrix and controllable SINR at the receiving end, A = [89.5 124.3 147.3 166]: (a) linear y axis; (b) log y axis.

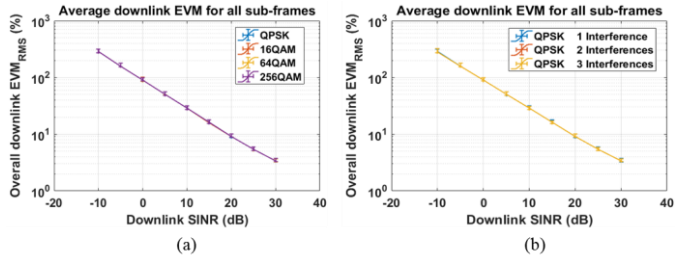

Figure 12: Normalised RMS EVM of the receiving signal interfered by in-band LTE CP-OFDM signals demodulated using previously obtained channel H-matrix and controllable SINR at the receiving end, A = 90: (a) comparisons of different symbol modulation schemes; (b) comparisons of different numbers of interferences.

Without adjusted SINR, the multipath environment will cause more uncertainty in the measurements. As shown in Figure 13. Note that, even though the payload data of each sub-fame is the same, the signal power of the sub-frames various due to different primary synchronisation symbol (PSS) allocations, nevertheless, a linear relationship can also be found between averages of SINR and RMS EVM.

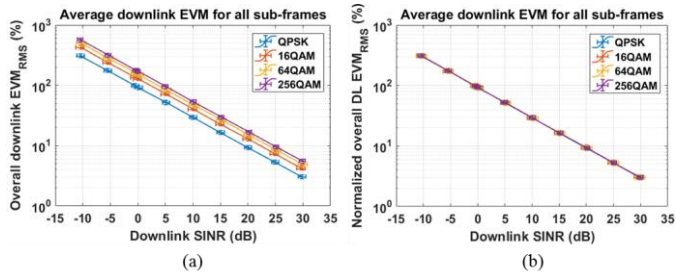

Figure 13: RMS EVM OTA measurement results of the receiving signal interfered by in-band GWN demodulated using previously obtained channel H-matrix: (a) un-normalised RMS EVM, A = [92.5 127.8 152.2 172.6]; (b) normalised RMS EVM, A = 92.5.

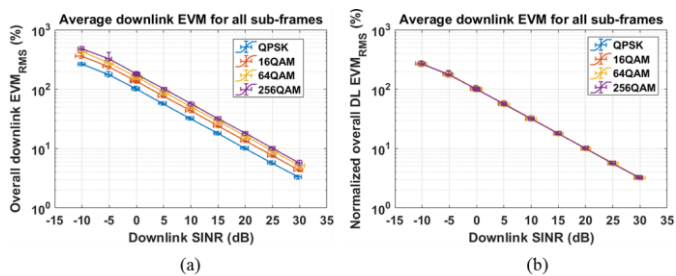

Figure 14: RMS EVM OTA measurement results for a received signal with by in-band GWN interference and demodulated using a channel H-matrix estimated in real-time: (a) un-normalised RMS EVM, A = [102.7 138 162.2 182.5] (SINR > 0)  (b) normalised RMS EVM, A = 102.7 (SINR > 0) .

Because the mobile cannot tell the interferences from the desired signals, when interference comes from other systems, the estimation of channel information will become inaccurate, which brings more uncertainties into the MIMO decoding process. As shown in Figure 14, the relationship between SINR and EVM appears to be nonlinear with the SINR increases.

## 4 Conclusion

A series of rigorous measurements for MIMO OTA characterisation with traceable SINR were presented in this paper. Work was carried out using a 4G base-station emulator and UE communication system and the NPL 5G MIMO testbed, including downlink and uplink EVM and throughput with respect to swept SINR. With respect to different measurement scenarios, measurement uncertainties of overall RMS EVM have been analysed. All the channel power measurements are reference to a traceable spectrum analyser and the results have been calibrated and corrected. The associated measurement uncertainties have been taken into account for obtaining traceable SINR.

## Acknowledgements

This work was supported by the National Science Foundation of China (under project number 61471193), the China Scholarship Council and the Funding of Jiangsu Innovation Program for Graduate Education (under project number KYLX15_0284). The work of T H Loh and D. Cheadle were supported by the 2017 – 2020 National Measurement System Programme of the UK government's Department for Business, Energy and Industrial Strategy (BEIS), under Science Theme Reference EMT17 of that Programme.

## References

[1] M. Rumney, *LTE and the Evolution to 4G Wireless: Design and Measurement Challenges*, 2nd Edition, Wiley-Blackwell, 2013.
[2] A. Osseiran, J. F. Monserrat. P. Marsch, *5G Mobile and Wireless Communications Technology*, Cambridge University Press, 2016.
[3] Jing, Ya, et al. "Overview of 5G UE OTA performance test challenges and methods", *IEEE MTT-S International Wireless Symposium (IWS)*, (2018).
[4] 3GPP TR 37.977 V14.6.0 (2018-07), "Verification of radiated multi-antenna reception performance of User Equipment (UE)", Jul. 2018.
[5] K. A. Hamdi. "On the statistics of signal-to-interference plus noise ratio in wireless communications", *Communication, IEEE Transactions on*, vol. 57, pp. 3199-3204, (2009).
[6] T. Brown, et al. "Prediction of SINR using BER and EVM for Massive MIMO Applications", *EuCAP 2018 Proceedings,* (2018).
[7] T. Brown, D. Humphreys, M. Hudlika, and M. Dieudonne. "Definition of SINR - MIMO, under 14IND10 MET5G A1.1.3", May 2017.
[8] Zhan, Zhiqiang, et al. "Measurement Uncertainty Evaluation of Digital Modulation Quality Parameters: Magnitude Error and Phase Error", *MATEC Web of Conferences*, vol. 61, EDP Sciences, (2016).